\documentclass[12pt]{article}

\usepackage[utf8]{inputenc}
\usepackage{amsmath, amssymb, amsthm, graphicx, tikz, xfrac}
\usepackage[capposition=top]{floatrow}
\usepackage[colorlinks=true, linkcolor=blue, citecolor=blue]{hyperref}
\usepackage{setspace}
\usepackage{natbib}
\usepackage{fullpage}
\usepackage{microtype}
\usepackage{bbm}
\usepackage{booktabs}
\usepackage{verbatim,color}
\usepackage{authblk}
\usepackage{color}

\DeclareMathOperator*{\argmax}{argmax}

\newcommand{\dd}{\,\mathrm{d}}

\title{Testing and Dating Structural Changes in Copula-based Dependence Measures\thanks{
Financial support by Deutsche Forschungsgemeinschaft (DFG grant ``Strukturbr\"uche und Zeitvariation in hochdimensionalen Abh\"angigkeitsstrukturen'') and the usage of the CHEOPS HPC cluster at the Regional Computing Center (RRZK) of the University of Cologne for parallel computing are gratefully acknowledged.}}

\author[1]{Florian Stark\thanks{Corresponding author: University of Cologne, Institute of Econometrics and Statistics, Meister-Ekkehart-Str. 9, 50937 Cologne, Germany. E: fstark3@uni-koeln.de. T: +49 221 470-4130. ORCID: 0000-0001-7419-6702.}}

\affil[1]{University of Cologne, Institute of Econometrics and Statistics}

\author[2]{Sven Otto}
\affil[2]{University of Bonn, Institute for Finance and Statistics}

\date{\today}

\usepackage{titlesec}
\titleformat{\section}[block]{\centering\normalfont\sffamily}{\thesection.}{0.5em}{\lsstyle\uppercase}
\titleformat{\subsection}[block]{\normalfont\sffamily}{\thesubsection.}{0.4em plus .1em minus .2em}{}
\titleformat{\subsubsection}[runin]{\normalfont\sffamily}{\thesubsubsection.}{0.4em plus .1em minus .2em}{}[.]
\titlespacing*\section{0pt}{18pt plus 4pt minus 2pt}{4pt plus 1pt minus 1pt}
\titlespacing*\subsection{0pt}{16pt plus 3pt minus 2pt}{4pt plus 1pt minus 1pt}
\titlespacing*\subsubsection{0pt}{12pt plus 2pt minus 1pt}{4pt plus 1pt minus 1pt}

\makeatletter
\def\mythanks#1{%
    \protected@xdef \@thanks {\@thanks \protect \footnotetext [\the \c@footnote ]{#1}}%
}
\makeatother

\begin{document}

{
\maketitle
\begin{abstract} 
This paper is concerned with testing and dating structural breaks in the dependence structure of multivariate time series.
We consider a cumulative sum (CUSUM) type test for constant copula-based dependence measures, such as Spearman's rank correlation and quantile dependencies.
The asymptotic null distribution is not known in closed form and critical values are estimated by an i.i.d.\ bootstrap procedure. 
We analyze size and power properties in a simulation study under different dependence measure settings, such as skewed and fat-tailed distributions.
To date break points and to decide whether two estimated break locations belong to the same break event, we propose a pivot confidence interval procedure. 
Finally, we apply the test to the historical data of ten large financial firms during the last financial crisis from 2002 to mid-2013. 
\end{abstract}
\noindent \textbf{JEL Classification:} C12, C14 \\
\noindent \textbf{Keywords:} Dependence measure testing, Spearman's rho, quantile dependence, portfolio optimization}

\newpage
\parindent0cm

\doublespacing

\section{Introduction}

The detection of structural breaks in statistics or statistical models is a broad research topic.
Early works can be found in \citet{page1954, page1955} and \cite{kiefer1959}, who were concerned with quality control problems. 
Tests for structural breaks in linear regression coefficients were proposed in \cite{Chow1960}, \cite{Brown1975}, \cite{Kramer1988}, \cite{Andrews1993}, and \cite{Bai1998}, among others.
Further prominent examples of change point analysis are the detection of instabilities in mean and variance (see \citealp{Horvath1999} and \citealp{aue2009a}).
For a review on recent developments see \cite{aue2013}.

A current research topic is analyzing changes in dependencies of financial variables such as stock returns. 
During the last financial crisis from 2007 onwards, it was observed that the dependence and volatility between financial market variables increased rapidly, which in turn led to inaccurate estimates and predictions of various risk figures (see \citealp{bissantz2011}).
Therefore, financial risk figures cannot be expected to remain constant over time (see \citealp{longin1995}). 
A portfolio manager is interested in reducing the risk and the amount of losses by dividing the assets into different investment opportunities. 
Such an effect is known as the diversification effect.
An increase in the dependence measures of asset returns can lead to the failure of portfolio diversification (see \citealp{sancetta2007}).
Structural break tests for dependence measures are therefore an important tool in portfolio management, as a detected break point indicates that the selected portfolio may no longer follow the previous correlation structure and that investment strategies should be adjusted.
Moreover, such tests, can be used to identify and quantify contagion between different financial markets.

The copula of random variables plays an important role in this context. 
\cite{schweizer1981} showed that any property of the joint distribution of two random variables that is invariant under strictly increasing transformations can be expressed as a function of their copula. 
Therefore, copula-based measures of dependence are of particular interest in practice. 
\cite{embrechts2002} argued that the copula provides the best understanding of the general concept of dependence for risk management. 
For an overview of copula and copula-based dependence measures see \cite{nelsen2006} and \cite{schmid2010}. 
In \cite{liebscher2014} the estimation of these measures is discussed. 
Parametric and semi-parametric approaches to testing for breaks in copula-based constraints are discussed in \cite{giacomini2009} and \cite{guegan2010}.
\citet{bucher2013} proposed a test for breaks in copula in the presence of general time dependencies. 
\cite{kutzker2019} considered a test for relevant changes in the copula. 
Recently, several tests for the constancy of certain dependence measures have also been developed. 
\citet{dehling2017} considered the case of Kendall's tau, \cite{wied2012} and \cite{posch2019} investigated the case of correlations, and \cite{wied2014} as well as \cite{kojadinovic2016} developed a test for the case of Spearman's rho.

In this paper we investigate a nonparametric test that was introduced in \cite{manner2019} as a test for detecting structural breaks in factor copula models.
\cite{manner2019} estimated the factor copula parameters using the simulated method of moments, while their nonparametric test statistic depends only on the copula-based moment conditions.
Therefore, we can apply their methodology also outside the framework of factor copula models, and extend their testing framework to the general problem of change point testing in cross-sectional dependencies of multivariate time series.
The test statistic is based on cumulative sums of a vector of different pre-specified dependence measures.
The measures are applied to residual data from pre-estimated marginal GARCH models so that the test is of nonparametric nature once we determined the residuals.
Under the null hypothesis there is no change in the dependence measures.
We focus primarily on Spearman's rho and quantile dependencies.
However, the dependence measure vector may in principle contain any measure that can be represented as a continuous function of the copula.

Since the asymptotic null distribution of the test statistic is not known in closed form and depends on the underlying joint distribution, we follow \cite{manner2019} and estimate the critical values by an i.i.d.\ bootstrap procedure.
To estimate break point locations and to identify equality of two estimated break points, we propose a heuristic procedure. 
For any break point estimate, we derive pivot confidence intervals using a percentile bootstrap procedure, and we consider two estimated break points as equal if they both lie in the intersection of their confidence intervals.
Moreover, the simulation studies in \cite{manner2019} are extended by analyzing size and power properties of the test for different skewed and fat tailed distributions for different settings of the used vector of dependence measures.
Finally, we provide a real data application on daily returns of ten large financial firms during the last financial crisis, in which we apply the test to the full period and to a rolling window of a fixed window size.

The paper is structured as follows. Section \ref{sec:theory} presents the test statistic, the break point estimator, and the confidence interval procedure. 
Results from Monte Carlo simulations can be found in Section \ref{sec:simulations}. 
Section \ref{sec:application} presents the empirical application, and Section \ref{sec:conclusion} concludes the paper.

\section{Testing for constancy in copula-based dependence measures}\label{sec:theory}

In this section we discuss the nonparametric test by \cite{manner2019} and its application to testing for change points in dependencies of multivariate time series.
The general heteroskedastic time series model is introduced in Section \ref{sec:model}, the hypothesis test is defined in Section \ref{sec:test}, and the estimation of break points and the identification of the equality of two estimated break points are discussed in \ref{sec:breakpointestimatiion}.

\subsection{The model} \label{sec:model}

We consider the semiparametric copula-based multivariate dynamic model, which was introduced in \cite{chen2006} and further studied in \cite{oh2013}, \cite{remillard2017}, and \cite{manner2019}.
Let $\pmb Y_t = (Y_{1,t}, \ldots, Y_{N,t})'$ be a multivariate time series with $t=1, \ldots, T$, and let $\mathcal F_t$ denote the sigma-algebra generated by $\{ \pmb Y_j, j \leq t\}$.
For each component $i=1, \ldots, N$, we assume that
\begin{equation*}
Y_{i,t} = \mu_{i,t}(\pmb \phi) + \sigma_{i,t}(\pmb \phi) \eta_{i,t},  \qquad t=1, \ldots, T,
\end{equation*}
where $\mu_{i,t}(\pmb \phi) = E[Y_{i,t}|\mathcal F_{t-1}]$ and $\sigma_{i,t}(\pmb \phi) = E[(Y_{i,t} - \mu_{i,t}(\pmb \phi))^2 | \mathcal F_{t-1}]$.
The error term $\eta_{i,t}$ has zero mean and unit variance by definition and is assumed to be identically distributed with continuous distribution function $F_i(x)$.
By Sklar's theorem, there exists a unique copula function $C_t$ such that the joint distribution function of the multivariate error term $\pmb \eta_t = (\eta_{1,t}, \ldots, \eta_{N,t})'$ is given by $F_{\eta,t}(x_1, \ldots, x_N) = C_t(F_1(x_1), \ldots, F_N(x_N))$.
The index $t$ indicates that cross-sectional dependence measures defined by the copula might not be constant over time. 
Note that $\pmb \eta_t$ is uncorrelated in the time domain, and its copula $C_t$ inherits the cross-sectional dependence structure of $\pmb Y_t$, making it particularly well suited for studying the dependencies in $\pmb Y_t$.

The parameter vector $\pmb \phi$ drives the dynamics of the conditional mean and variance, and we assume to have a $\sqrt T$ consistent estimator $\hat{\pmb \phi}$ for $\pmb \phi$, which holds true for ARMA and GARCH models under fairly mild conditions.
Let $\hat \eta_{i,t} = \sigma_{i,t}^{-1}(\hat{\pmb \phi}) (Y_{i,t} - \mu_{i,t}(\hat{\pmb \phi}))$ be the residual data,  
and let $\hat{\pmb \eta}_t = (\hat \eta_{1,t}, \ldots, \hat \eta_{N,t})'$ denote the vector of residuals.

\subsection{Testing problem} \label{sec:test}

The test proposed by \cite{manner2019} compares sequentially estimated dependence measure vectors to the full sample estimated analogue using a CUSUM procedure.
While their method is designed to test the null hypothesis of constant factor copula parameters in the equidependence model setting of \cite{oh2017}, we can apply their test to the more general hypothesis of constant cross-sectional dependencies in multivariate time series.

Let $C_t^{ij}$ denote the bivariate marginal copula of $C_t$ that corresponds to the $i$-th and $j$-th margins, and let $m_t^{ij}$ be a vector of pairwise dependence measures of the variables $i$ and $j$ at time $t$.
The dependence measures in the vector $m_t^{ij}$ are pre-specified in advance and may consist of any statistic that can be expressed as a continuous function of $C_t^{ij}$.
An overview of suitable measures can be found in \cite{schmid2010}.
As in the work of \cite{oh2013}, we focus on Spearman's rank correlation $\rho_t^{ij}$ and quantile dependence measures $\lambda_{q,t}^{ij}$. These can be defined in terms of the copula as
\begin{align}
\rho_t^{ij}:=&12\int_0^1\int_0^1 C_t^{ij}(u,v) \dd u \dd v-3, \label{eq:rho} \\
\lambda_{q,t}^{ij}:=&\begin{cases}
q^{-1} C_t^{ij}(q,q),  &q\in(0,0.5],\\
(1-q)^{-1} ( 1-2q+C_t^{ij}(q,q)), &q\in(0.5,1). \label{eq:lambda}
\end{cases}
\end{align}
Under the null hypothesis, the copula $C_t$ is time constant with constant pairwise dependence measure vectors,
\[
H_0: m_1^{ij}=m_2^{ij}=\dots=m_T^{ij} \quad \forall i,j \in 
\{1, \ldots, N\}
, \ i \neq j.
\]
Under the alternative, there is single break point at an unknown time $t \in \lbrace 1,\dots,T-1\rbrace$, such that
\begin{equation*}
H_1: m_{1}^{ij} = \ldots = m_{t}^{ij}\neq m_{t+1}^{ij} = \ldots = m_{T}^{ij} \qquad \text{for some } i,j \in \{1, \ldots, N\}, i \neq j.
\end{equation*}

For any $t \in \{1, \ldots, T\}$ and $i,j \in \{1, \ldots N\}$, the sequential empirical distribution function and the sequential empirical bivariate copula are given as
\[
\hat{F}_{i,t}(y):=\frac{1}{t}\sum\limits_{k=1}^{t}\mathbbm{1}\lbrace \hat{\eta}_{ik}\leq y\rbrace, \quad
\hat{C}_t^{ij}(u,v):=\frac{1}{t}\sum\limits_{k=1}^{t}\mathbbm{1}\lbrace \hat{F}_{i,t}(\hat{\eta}_{ik})\leq u,\hat{F}_{j,t}(\hat{\eta}_{jk})\leq v\rbrace.
\]
The sequential sample counterparts of the dependence measures in \eqref{eq:rho} and \eqref{eq:lambda} are defined as
\begin{align*}
\hat{\rho}^{ij}_t:=&\frac{12}{t}\sum_{k=1}^{t}\hat{F}_{i,t}(\hat{\eta}_{ik})\hat{F}_{j,t}(\hat{\eta}_{jk})-3,\\
\hat{\lambda}_{q,t}^{ij}:=&\begin{cases}
q^{-1} \hat{C}_t^{ij}(q,q), &q\in(0,0.5],\\
(1-q)^{-1} (1-2q+\hat{C}_t^{ij}(q,q)),  &q\in(0.5,1),
\end{cases}
\end{align*}
and the consistency of the empirical copula-based measures is discussed in \cite{liebscher2014}.
Moreover, the sequential sample analogue of the dependence measure vector $m_t^{ij}$ is denoted as $\hat{m}_{t}^{ij}$, and, following the equidependence setting of \cite{oh2017} and \cite{manner2019}, we consider the pairwise averaged sequential dependence measure vectors given by
\begin{equation*}
	\hat{m}_{t} = \frac{2}{N(N-1)} \sum_{i=1}^{N-1} \sum_{j=i+1}^N \hat{m}_{t}^{ij}, \qquad t=1, \ldots, T.
\end{equation*}
The CUSUM-type test statistic is based on the maximum difference between the recursive estimates and the full sample estimate of the dependence measure vector. Formally, it is defined as
\begin{equation*}
M_T:= \max_{\varepsilon T \leq t \leq T} \left(\frac{t}{T}\right)^2 T (\hat{m}_t-\hat{m}_T)'(\hat{m}_t-\hat{m}_T).
\end{equation*}
In \cite{manner2019} it is noted that the trimming parameter $\varepsilon$ has to be chosen strictly greater than zero, and in practice it should be chosen in a way that we have enough data information to receive reasonable dependence measure vector estimates. Throughout the rest of the paper, we use $\varepsilon=0.1$.

The test rejects the null hypothesis if $M_T>q_{1-\alpha}$,
where $q_{1-\alpha}$ is the $(1-\alpha)$-quantile of the limiting distribution of $M_T$, and $\alpha$ is the significance level.
\cite{manner2019} imposed regularity assumptions on the underlying copula $C$, which ensure that the estimated rank correlation and quantile dependencies converge to their respective population counterparts (see Assumptions 1 and 2 in \citealp{manner2019}).
Under these regularity assumptions and the null hypothesis, Lemma 7 in \cite{manner2019} implies that
\begin{equation}
M_T \overset{d}{\longrightarrow}\underset{s\in[\varepsilon,1]}{\textsf{sup}}(A(s)-sA(1))'(A(s)-sA(1)), \label{teststat}
\end{equation}
as $T \to \infty$,
where $A(s)$ is some Gaussian process.
Without further assumptions on the copula, the covariance structure of $A(s)$ and the limiting distribution of $M_T$ are unknown.
Therefore, critical values cannot be computed or simulated directly. 
To overcome this issue, a bootstrap procedure similar to the one in \cite{manner2019} is considered:
\begin{itemize}
\item[i)] For $p=1,\dots, B$, sample with replacement from $\lbrace\hat{\eta}_t\rbrace_{t=1}^T$ to obtain $\lbrace\hat{\eta}_t^{(p)}\rbrace_{t=1}^T$.
\item[ii)] For $p=1,\dots, B$ and $t=\varepsilon T, \dots, T$, compute $\hat{m}_{t}^{(p)}$ from $\lbrace\hat{\eta}_k^{(p)}\rbrace_{k=1}^t$ and $\hat{m}_T$ from $\lbrace\hat{\eta}_t\rbrace_{t=1}^T$.
\item[iii)] For $p=1,\dots, B$, let $A^{(p)}(t/T):= \frac{t}{T} \sqrt{T} (\hat{m}_t^{(p)}-\hat{m}_T)$, and calculate the bootstrap analogue of \eqref{teststat} given by $K^{(p)}:= \max_{\{\varepsilon T \leq t \leq T\rbrace} (A^{(p)}(t/T)-\frac{t}{T}A^{(p)}(1))'(A^{(p)}(t/T)-\frac{t}{T}A^{(p)}(1))$.
\item[iv)] Determine the bootstrap critical value $\hat q_{1-\alpha}$ such that $B^{-1} \sum_{p=1}^B \mathbbm{1}\lbrace K^{(p)}>\hat q_{1-\alpha} \rbrace = \alpha$.
\end{itemize}
The validity of this bootstrap procedure is discussed in \cite{manner2019}.

\subsection{Estimation of break points} \label{sec:breakpointestimatiion}

If we reject the null hypothesis, we speak of a structural break.
The estimation of the change point location, once we detected a structural break, is embedded in calculating the test statistic and is given by $\hat{k}:=\lfloor \hat{s}T\rfloor$, where
\begin{align}\label{bp}
\hat{s}=
\argmax_{\varepsilon T \leq t \leq T} \left(\frac{t}{T}\right)^2 T (\hat{m}_t-\hat{m}_T)'(\hat{m}_t-\hat{m}_T).
\end{align}

For a better comparison of estimated break point locations in practice that are determined by different dependence measure settings, we propose a heuristic procedure. 
This allows us to make a statement about whether two estimated break point locations $\hat{s}_a$ and $\hat{s}_b$, $a\neq b$, belong to the same class of break points.
The subscripts $a$ and $b$ denote the choice of a different vector of dependence measures, $\hat{m}^{ij}_{t,a}$ and $\hat{m}^{ij}_{t,b}$.
Note that the break point location estimator defined in equation \eqref{bp} is a scalar in the uniform interval $(0,1]$. 
We define pivot confidence intervals $\hat{K}_a:=[\hat{K}_a^-,\hat{K}_a^+]:=[2\hat{s}_a-\hat{c}^a_{1-\frac{\alpha}{2}},2\hat{s}_a -\hat{c}^a_{\frac{\alpha}{2}}]$ and $\hat{K}_b:=[2\hat{s}_b-\hat{c}^b_{1-\frac{\alpha}{2}},2\hat{s}_b -\hat{c}^b_{\frac{\alpha}{2}}]$, where $\hat{c}^a_{(\cdot)}$ and $\hat{c}^b_{(\cdot)}$ are estimated quantiles of the bootstrap distribution of $\hat{s}_a$ and $\hat{s}_b$, which can be determined by using a percentile bootstrap procedure.
Suppose that we have detected two break point locations $\hat{s}_a$ and $\hat{s}_b$ when using the dependence measures $\hat{m}^{ij}_{t,a}$ and $\hat{m}^{ij}_{t,b}$.
The bootstrap is defined as follows:
\begin{itemize}
\item[i)] For $\hat{m}^{ij}_{t,a}$, split the sample into $\lbrace \hat{\pmb{\eta}}_t\rbrace_{t=1}^{\lfloor \hat{s}_aT \rfloor}$ and $\lbrace \hat{\pmb{\eta}}_t\rbrace_{t=\lfloor \hat{s}_aT \rfloor +1}^T$, and for $\hat{m}^{ij}_{t,b}$, split into $\lbrace \hat{\pmb{\eta}}_t\rbrace_{t=1}^{\lfloor \hat{s}_bT \rfloor}$ and $\lbrace \hat{\pmb{\eta}}_t\rbrace_{t=\lfloor \hat{s}_bT \rfloor +1}^T$.

\item[ii)] For $p=1,\dots, B$, sample with replacement from $\lbrace \hat{\pmb{\eta}}
_t\rbrace_{t=1}^{\lfloor \hat{s}_aT\rfloor }$ and $\lbrace \hat{\pmb{\eta}}_t\rbrace_{t=\lfloor \hat{s}_aT \rfloor +1}^T$ to obtain $\lbrace \hat{\pmb{\eta}}^{(p)}_{t,a}\rbrace_{t=1}^{T}$, and from $\lbrace \hat{\pmb{\eta}}_t\rbrace_{t=1}^{\lfloor \hat{s}_bT \rfloor }$ and $\lbrace \hat{\pmb{\eta}}_t\rbrace_{t=\lfloor \hat{s}_bT \rfloor +1}^T$ to obtain $\lbrace \hat{\pmb{\eta}}^{(p)}_{t,b}\rbrace_{t=1}^{T}$.

\item[iii)] For $p=1,\dots, B$, estimate $\hat{s}_a^{(p)}$ and $\hat{s}_b^{(p)}$ from $\lbrace \hat{\pmb{\eta}}^{(p)}_{t,a}\rbrace_{t=1}^{T}$ and $\lbrace \hat{\pmb{\eta}}^{(p)}_{t,b}\rbrace_{t=1}^{T}$ using \eqref{bp}.

\item[iv)] Compute sample quantiles $\hat{c}^a_{\frac{\alpha}{2}}, \hat{c}^b_{\frac{\alpha}{2}}$ and $\hat{c}^a_{1-\frac{\alpha}{2}}, \hat{c}^b_{1-\frac{\alpha}{2}}$ from $\lbrace\hat{s}_a^{(p)}\rbrace_{p=1}^B$ and $\lbrace\hat{s}_b^{(p)}\rbrace_{p=1}^B$.
\end{itemize}

We consider two estimated break point locations $\hat{s}_a$ and $\hat{s}_b$ as being originated from the same break if both are in the intersection of the two confidence intervals, i.e. if $\hat{s}_a,\hat{s}_b\in \hat{K}_a\cap \hat{K}_b$.
Note that this procedure is only plausible if we consider the same testing period for both dependence settings $\hat{m}^{ij}_{t,a}$ and $\hat{m}^{ij}_{t,b}$.  
Therefore, in the empirical application the procedure can only be applied for a break comparison in the full sample testing and cannot be used in the rolling window testing procedure, since similar break point locations may belong to different tested periods.
Furthermore, notice that the estimation error between the estimated break point $\hat{s}$ and the true break point $s_0$ is approximately the same as the difference between the estimated break location $\hat{s}$ and the bootstrap break estimate $\hat{s}^{(p)}$, i.e.
\begin{align*}
1-\alpha&\approx P(\hat{c}_{\frac{\alpha}{2}}\leq \hat{s}^{(p)}\leq \hat{c}_{1-\frac{\alpha}{2}})\\
&=P(\hat{k}-\hat{c}_{1-\frac{\alpha}{2}}\leq \hat{s}-\hat{s}^{(p)}\leq \hat{s}-\hat{c}_{\frac{\alpha}{2}})\\
&\approx P(\hat{s}-\hat{c}_{1-\frac{\alpha}{2}}\leq s_0-\hat{s}\leq \hat{s}-\hat{c}_{\frac{\alpha}{2}})\\
&= P(2\hat{s}-\hat{c}_{1-\frac{\alpha}{2}}\leq s_0\leq 2\hat{s}-\hat{c}_{\frac{\alpha}{2}}).
\end{align*}
A similar procedure for a different change point test setting was considered in \cite{huskova2008}.

\section{Simulations}\label{sec:simulations}
In this section we analyze size and power properties of the proposed test and the validity of the confidence interval procedure for different dependence measure settings in Monte Carlo simulations.
We consider the following dependence measure settings:
\begin{align*}
\hat{m}_{1,t}^{ij}&=\big(\hat{\rho}_t^{ij}, \hat{\lambda}_{0.05,t}^{ij}, \hat{\lambda}_{0.1,t}^{ij}, \hat{\lambda}_{0.9,t}^{ij}, \hat{\lambda}_{0.95,t}^{ij}\big)',\\
\hat{m}_{2,t}^{ij}&=\big( \hat{\lambda}_{0.05,t}^{ij}, \hat{\lambda}_{0.1,t}^{ij}, \hat{\lambda}_{0.9,t}^{ij}, \hat{\lambda}_{0.95,t}^{ij}\big)',\\
\hat{m}_{3,t}^{ij}&=\big( \hat{\lambda}_{0.9,t}^{ij}, \hat{\lambda}_{0.95,t}^{ij}\big)',\\
\hat{m}_{4,t}^{ij}&=\big( \hat{\lambda}_{0.05,t}^{ij}, \hat{\lambda}_{0.1,t}^{ij}\big)',\\
\hat{m}_{5,t}^{ij}&=\hat{\rho}_t^{ij}.
\end{align*}

The measures are applied to the residual process $\{\pmb \eta_t\}_{t=1}^T$, which is simulated under three different copula models, including skewed and fat tailed distributions.
For all simulations, we consider a significance level of $\alpha = 5\%$ and 301 Monte Carlo repetitions. 
Due to the fact that we are mainly interested in comparing different dependence settings, we consider the fixed time and cross-sectional dimensions $T=1000$ and $N=10$.  
An analysis for different combinations of $T$ and $N$ for the setting $\hat{m}_1^{ij}$ under a factor copula can be found in \cite{manner2019}.

First, we consider a simple one factor copula model following \citet{oh2013, oh2017}.
The copula is implied by the factor structure
\begin{equation}
	\eta_{it} = \theta_{t} Z + q_{it} \label{3.1}, \quad i= 1,\dots,N, \ t=1,\dots,T,
\end{equation}
where $Z\overset{}{\sim}\mathsf{Skew\ t}\left(\nu^{-1},\lambda\right)$, which refers to the skewed t-distribution by \cite{hansen1994}, and $q_{it} \overset{i.i.d.}{\sim}\mathsf{t}\left(\nu^{-1}\right)$.
We fix $\nu^{-1}=0.25$ and consider $\lambda \in \lbrace-0.5, 0, 0.5\rbrace$.
For the time varying parameter vector $\theta_t$, we consider a single break point at $t=T/2$, where $\theta_t=\theta^0 = 1$ for $t=1, \ldots, T/2$, and $\theta_t = \theta^{1} \in \lbrace 1, 1.1, 1.2, 1.3, 1.4, 1.5\rbrace$ for $t= T/2+1, \ldots, T$.

\begin{table}[t!]\caption{Size and power under the factor copula model}\label{power1}
\footnotesize
  \begin{center}
    \begin{tabular}{r|rrrrrrr}
    \toprule
      $T=1000,N=10$ &  $\theta^1=1$     & $\theta^1=1.1$    & $\theta^1=1.2$& $\theta^1=1.3$   & $\theta^1=1.4$     & $\theta^1=1.5$\\
    \midrule
        &\multicolumn{5}{c}{$\lambda=-0.5$} \\
        \midrule
      $\hat{m}_{1,t}^{ij}$
                  &0.0465     &0.1628   &0.3887   &  0.6013   &  0.8007   &0.9269  \\
              
                  \midrule
     $\hat{m}_{2,t}^{ij}$
      
                  &0.0498     &0.1462   &0.3189   &   0.5249  &  0.6944   &0.8704  \\
            
                  \midrule
   $\hat{m}_{3,t}^{ij}$
               
                    &0.0365     &0.0897  &0.2259   &  0.4784   &  0.7043   & 0.8571 \\
                
    \midrule
     $\hat{m}_{4,t}^{ij}$
                 
                  &0.0498     &0.1329  &0.2724   &   0.4286  &   0.5781  & 0.7176 \\
                
                    \midrule
    $\hat{m}_{5,t}^{ij}$
                  
                   &0.0532     &0.2558   &0.6645   &  0.9435   &  0.9934   &1.0000 \\
\midrule
    &\multicolumn{5}{c}{$\lambda=0$} \\
    \midrule
    
     $\hat{m}_{1,t}^{ij}$
 
                  &0.0532    &0.1993   &0.4485  &  0.7010   &  0.9003  &0.9767 \\
              
                  \midrule
     $\hat{m}_{2,t}^{ij}$
      
                  &0.0532    &0.1927   &0.3787  &  0.6213   &  0.8538  &0.9358 \\
            
                  \midrule
     $\hat{m}_{3,t}^{ij}$
               
                   &0.0432    &0.1229   &0.2625  &  0.4385   &  0.6478  &0.8206 \\
                
    \midrule
      $\hat{m}_{4,t}^{ij}$
                 
                &0.0565   &0.1495   &0.2857 &  0.4485   &  0.6146  &0.8641  \\
                
                    \midrule
     $\hat{m}_{5,t}^{ij}$
                  
                  &0.0598    &0.2791   &0.7176  &  0.9668   &  0.9967  &1.0000 \\
                  \midrule
             &\multicolumn{5}{c}{$\lambda=0.5$} \\ 
             \midrule 
      $\hat{m}_{1,t}^{ij}$
 
                  &0.0565    &0.1661   &0.3322  &  0.5781   &  0.8538  &0.9635  \\
              
                  \midrule
     $\hat{m}_{2,t}^{ij}$
      
                  &0.0764    &0.1495   &0.2890  &  0.4917   &  0.7342  &0.9203 \\
            
                  \midrule
     $\hat{m}_{3,t}^{ij}$
               
                   &0.0731    &0.1395   &0.2658  &  0.3854   &  0.5781  &0.7741 \\
                
    \midrule
      $\hat{m}_{4,t}^{ij}$
                 
                &0.0332    &0.1096   &0.2558  &  0.4651   &  0.6611  &0.8538  \\
                
                    \midrule
     $\hat{m}_{5,t}^{ij}$
                  
                  &0.0498    &0.2658   &0.7043  &  0.9502   &  1.0000  &1.0000 \\
                  \bottomrule
    \end{tabular}
    \end{center} 
   \parbox{13cm}{ \vspace{-1.5ex} 
\scriptsize
Note: Rejection rates for different break sizes and different measure combinations simulated under the model \eqref{3.1} are reported. Since $\theta^0=1$, the first column refers to the case of no break.
}
\end{table}

The size and power results are presented in Table \ref{power1}. 
The highest power is obtained for $\hat{m}_5^{ij}$, followed by the setting $\hat{m}_1^{ij}$.
The cases where only upper or lower quantiles are considered exhibit poor power properties compared to the other dependence settings.
Considering both upper and lower quantile dependencies yields better power properties than if the cases are considered individually.

Furthermore, we simulate residual data from a Clayton copula and a Gumbel copula.
Note that Clayton and Gumbel copulas produce different tail dependence. 
Under the Clayton copula we have strong lower quantile dependence and the Gumbel copula produces strong upper quantile dependence.
For the Clayton copula, we consider the pre-break parameter $\theta^0=2.5$ and the post-break parameters $\theta^1=\lbrace 2.5,  3.0, 3.5, 4.0, 5.0,5.5\rbrace$, whereas, for the Gumbel copula, we set $\theta^0=2.0$ and $\theta^1=\lbrace 2.0, 2.2, 2.4, 2.6, 2.8, 3.0\rbrace$, where, as above, the break point is located in the middle of the sample.
Note that the parameters $\theta^0$ and $\theta^1$ are chosen so that the implied upper quantile dependence for the Clayton copula and implied lower quantile dependence for the Gumbel copula are of the same magnitude.

\begin{table}[t]\caption{Size and power under the Clayton copula}\label{power4}
\footnotesize
  \begin{center}
    \begin{tabular}{r|rrrrrrr}
    \toprule
      $T=1000,N=10$&  $\theta^1=2.5$     & $\theta^1=3.0$    & $\theta^1=3.5$& $\theta^1=4.0$   & $\theta^1=4.5$     & $\theta^1=5.0$\\
    \midrule
      $\hat{m}_{1,t}^{ij}$
 
                  &0.0532    &0.1960   &0.4518  &  0.7342   &  0.9336 &0.9701 \\
              
                  \midrule
    $\hat{m}_{2,t}^{ij}$
      
                   &0.0565    &0.1761  &0.3488  & 0.6179   &  0.8372 &0.8970 \\
            
                  \midrule
    $\hat{m}_{3,t}^{ij}$
               
                   &0.0631   &0.1827  &0.4219  &  0.7010   &  0.8738 &0.9402 \\
                
    \midrule
     $\hat{m}_{4,t}^{ij}$
                 
                &0.0565   &0.1694  &0.2924  &  0.3821   &  0.5382 &0.6047 \\
                
                    \midrule
    $\hat{m}_{5,t}^{ij}$
                  
                   &0.0332    &0.3854  &0.9468  &  1.0000   &  1.0000 &1.0000 \\

    \bottomrule
    \end{tabular}
    \end{center} 
      \parbox{13.5cm}{ \vspace{-1.5ex} 
\scriptsize
Note: Rejection rates for different break sizes and different measure combinations under the Clayton copula model are presented. Since $\theta^0=2.5$, the first column refers to the case of no break.
}
\end{table}%

\begin{table}[t]\caption{Size and power under the Gumbel copula}\label{power5}
\footnotesize
  \begin{center}
    \begin{tabular}{r|rrrrrrr}
    \toprule
      $T=1000,N=10$&  $\theta^1=2.0$     & $\theta^1=2.2$    & $\theta^1=2.4$& $\theta^1=2.6$   & $\theta^1=2.8$     & $\theta^1=3.0$\\
    \midrule
      $\hat{m}_{1,t}^{ij}$
 
                  &0.0399    &0.1628   &0.4352  &  0.8671   &  0.9668 &1.0000 \\
              
                  \midrule
     $\hat{m}_{2,t}^{ij}$
      
                   &0.0365  &0.1329   &0.3654  &  0.7874   &  0.9003 &0.9834 \\
            
                  \midrule
     $\hat{m}_{3,t}^{ij}$
               
                   &0.0399    &0.1229   &0.2492  & 0.4618   &  0.5648 &0.6678 \\
                
    \midrule
      $\hat{m}_{4,t}^{ij}$
                 
                &0.0565   &0.3522   &0.6445  &  0.8571   &  0.9402 &0.9734 \\
                
                    \midrule
     $\hat{m}_{5,t}^{ij}$
                  
                   &0.0532    &0.5282  &0.9435  &  1.0000   &  1.0000 &1.0000 \\
    \bottomrule
    \end{tabular}
    \end{center} 
      \parbox{13.5cm}{ \vspace{-1.5ex} 
\scriptsize
Note: Rejection rates for different break sizes and different measure combinations under the Gumbel copula model are presented. Since $\theta^0=2$, the first column refers to the case of no break.
}
\end{table}

\begin{center}
\begin{figure}[p]
\centering
\includegraphics[width=0.72\textwidth]{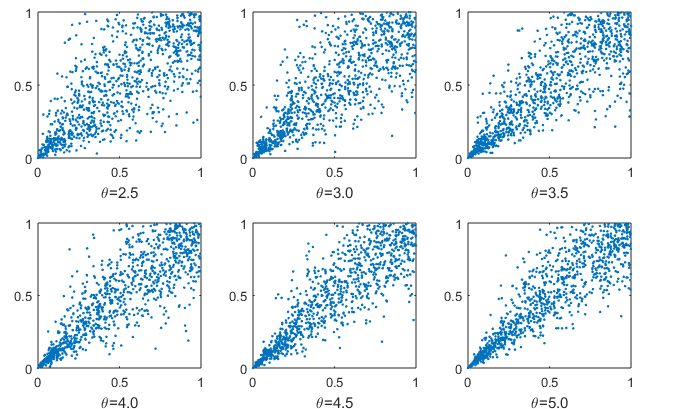}
\caption{Simulated points $(u,v)$, using a Clayton copula with $T=1000$ and fixed $\theta$. }
\label{clayton_pow}
\end{figure}
\end{center}

\begin{center}
\begin{figure}[p]
\centering
\includegraphics[width=0.65\textwidth]{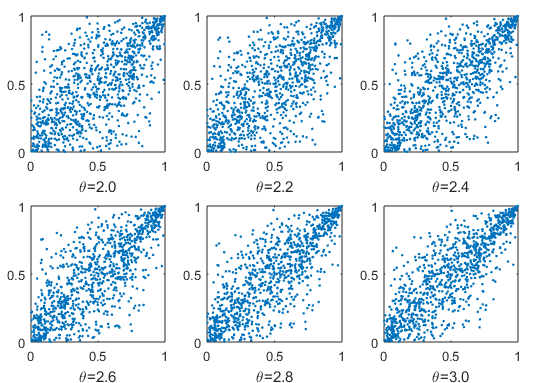}
\caption{Simulated points $(u,v)$, using a Gumbel copula with $T=1000$ and fixed $\theta$. }
\label{gumbel_pow}
\end{figure}
\end{center}

The size and power results are presented in Tables \ref{power4} and \ref{power5}.
Figures \ref{clayton_pow} and \ref{gumbel_pow} show the heavy tailed characteristics of the Clayton and the Gumbel copula for different post-break parameter values.
Note that the dependence structure in the lower (Clayton) and upper (Gumbel) cases just changes slightly, 
and, consequently, the test yields poor power properties in the cases $\hat{m}_4^{ij}$ and $\hat{m}_3^{ij}$, respectively.
Similar results are obtained for the factor copula model in the case of $\lambda \in \lbrace -0.5, 0.5\rbrace$ if only upper or lower quantile dependencies are used (see Table \ref{power1}).
Therefore, a combination of lower and upper quantile dependence measures as in $\hat{m}_2^{ij}$ yields better power properties. 
As in the case of the factor copula, the dependence vector settings $\hat{m}_1^{ij}$ and $\hat{m}_5^{ij}$ provide the best power properties of the test. 
This can be explained due to the fact that quantile dependencies suffer from a small number of data points in the tails, since, for the sample size of $T=1000$, we consider only 100 or 50 data points when computing the $\alpha$ and $(1-\alpha)$ quantiles, respectively.  
Consequently, for the quantile dependencies, a larger sample size is required to obtain the same power properties as in the case of Spearman's rho, where the rank correlation coefficient is computed from the entire sample and is a global dependence measure.

Although it would be plausible that more dependence measures within the dependence measure vector yield higher rejection rates of the test, this is not the case for the setting $\hat{m}_1^{ij}$.
The results indicate that a dependence vector, which is composed of a selection of quantile dependencies and the rank correlation, inherits the poor performance characteristics of the quantile dependencies, and that the better performance compared to the settings $\hat{m}_2^{ij},\hat{m}_3^{ij}$ and $\hat{m}_4^{ij}$ is mainly due to the fact that the rank correlation coefficient is used.

On the other hand, the use of various dependence settings may provide different break point estimates (see the empirical application in Section \ref{sec:application}). 
In time periods of a pronounced structural break, e.g.\ in periods in which extraordinary events on the financial market can be attributed, one can be more certain whether the detected break is plausible if several dependence settings lead to the same break event.
Another possibility is to divide the data into suitable subgroups and test these separately for structural breaks by using different combinations of the considered dependence measures. 
For example, the data could be divided into different industrial sectors. 
To test for the equality of two identified break points, we can use the confidence interval procedure defined in Section \ref{sec:breakpointestimatiion}.

In what follows, we present a small simulation study for the confidence interval procedure, where we use the dependence measure settings $\hat{m}_1^{ij}$ and $\hat{m}_3^{ij}$.
We simulate $\eta_{it}$ according to the DGP in \eqref{3.1} with $\lambda=-0.5$ and a single break in $\theta_t$ at $t=T/2$, i.e. $s_0=0.5$.
We fix the cross-sectional dimension at $N=10$ and consider different sample sizes of $T \in \lbrace 500, 1000, 1500\rbrace$ and break sizes of $\theta^1 \in \lbrace 1.5, 2.0, 2.5\rbrace$. 
For all simulations we use $B=500$ bootstrap replications.
In Table \ref{cover} we present the coverage probabilities $P(0.5\in\hat{K}_1)$ and $P(0.5\in\hat{K}_3)$, and the probability that the constructed break at $s_0$ lies in the intersection of $\hat{K}_1\cap\hat{K}_3$, i.e. $P(0.5\in\hat{K}_1\cap\hat{K}_3)$.
The coverage probability of $\hat{K}_1$ and $\hat{K}_3$ tends to $1-\alpha=0.95$ with increasing sample and break size.
The probability that the actual break at $s_0$ lies in the interval $\hat{K}_1\cap\hat{K}_3$ tends to $(1-\alpha)^2$, as $T \to \infty$.
Note, that in practice the size level $\alpha^*$ of the common break test can be controlled by considering $(1-\alpha)^2=1-\alpha^*$.

\begin{table}[t]\caption{Coverage probabilities under a single break point setting}\label{cover}
\footnotesize
  \begin{center}
    \begin{tabular}{r|rrr}
    \toprule
     &  \multicolumn{3}{c}{$\big(P(0.5\in\hat{K}_1)\ P(0.5\in\hat{K}_3)\ P(0.5\in\hat{K}_1\cap\hat{K}_3)\big)$}\\
    \midrule
     $B=500,N=10$& $T=500$     & $T=1000$    & $T=1500$\\
    \midrule
      $\theta_1=1.5$&(0.80\ 0.78\ 0.64)    &(0.93\ 0.87\ 0.81)  &(0.94\ 0.90\ 0.84)  \\
    \midrule
     $\theta_1=2.0$&(0.91\ 0.88\ 0.80)&(0.95\ 0.93\ 0.89)  &(0.95\ 0.92\ 0.89)\\
    \midrule
     $\theta_1=2.5$&(0.93\ 0.94\ 0.88)  &(0.93\ 0.94\ 0.88)   &(0.95\ 0.95\ 0.91)\\
    \bottomrule
    \end{tabular}
    \end{center} 
      \parbox{12cm}{ \vspace{-1.5ex} 
\scriptsize
Note: Coverage probabilities of confidence intervals $\hat{K}_1$ and $\hat{K}_3$ for a break at $0.5$ and the coverage probability of $0.5\in \hat{K}_1\cap \hat{K}_3$ are reported, where the data is simulated under the factor copula model \eqref{3.1}
}
\end{table}%

Under the same setting as above, we also simulate a case of two break points given by two residual data sets $\{\eta_{it}^{(1)}\}_{t=1}^T$ and $\{\eta_{it}^{(2)}\}_{t=1}^T$, where the breaks are located at $s_0^{(1)}=6/14=0.429$ for the first set and $s_0^{(2)}=0.5$ for the second set.
Note that this simulation setting mimics the situation where we split our sample in subsets.
The results are presented in Table \ref{cover1}.

\begin{table}[t]\caption{Coverage probabilities under a setting with two simulated break points}\label{cover1}
\footnotesize
  \begin{center}
    \begin{tabular}{r|ccc}
    \toprule
     &  \multicolumn{3}{c}{$\big(P(0.429\in\hat{K}_1)\ P(0.5\in\hat{K}_3)\ P(0.429, 0.5\in\hat{K}_1\cap\hat{K}_3)\big)$}\\
    \midrule
     $B=500,N=10$& $T=500$     & $T=1000$    & $T=1500$\\
    \midrule
      $\theta_1=1.5$&(0.81\ 0.78\ 0.46)    &(0.91\ 0.87\ 0.52)  &(0.92\ 0.90\ 0.31) \\
    \midrule
     $\theta_1=2.0$&(0.93\ 0.88\ 0.21)    &(0.95\ 0.93\ 0.01)  &(0.96\ 0.92\ 0.00)\\
    \midrule
     $\theta_1=2.5$&(0.95\ 0.94\ 0.01)    &(0.95\ 0.94\ 0.00)  &(0.95\ 0.95\ 0.00)\\
    \bottomrule
    \end{tabular}
    \end{center} 
      \parbox{12cm}{ \vspace{-1.5ex} 
\scriptsize
Note: Coverage probabilities of confidence intervals $\hat{K}_1$ and $\hat{K}_3$ for a breaks constructed at $0.5$ and $0.429$ and the coverage probability of $0.429, 0.5\in \hat{K}_1\cap \hat{K}_3$ are reported, where the data is simulated under the factor copula model \eqref{3.1}
}
\end{table}

The coverage probabilities of $\hat{K}_1$ for a break at $0.429$ and $\hat{K}_3$ for a break at $0.5$ tend to $1-\alpha=0.95$ with increasing sample and break size, whereas the results for $P(0.5\in\hat{K_3})$ are the same as those in Table \ref{cover}.
On the other hand, the probability that the two break points $0.429$ and $0.5$ lie in the interval $\hat{K}_1\cap\hat{K}_3$ tends to zero with increasing sample break size.
For example, a break step of $\theta^1=1.5$ implies a rank correlation change before and after the break of 0.17, whereas a break change of $\theta^1=2.0$ implies a rank correlation change of 0.25.
Thus, we conclude that the procedure is reasonably sized and has good power properties if the break steps and the sample size are high enough.

\section{Application}\label{sec:application}

\begin{figure}[t!]
\includegraphics[width=0.9\textwidth]{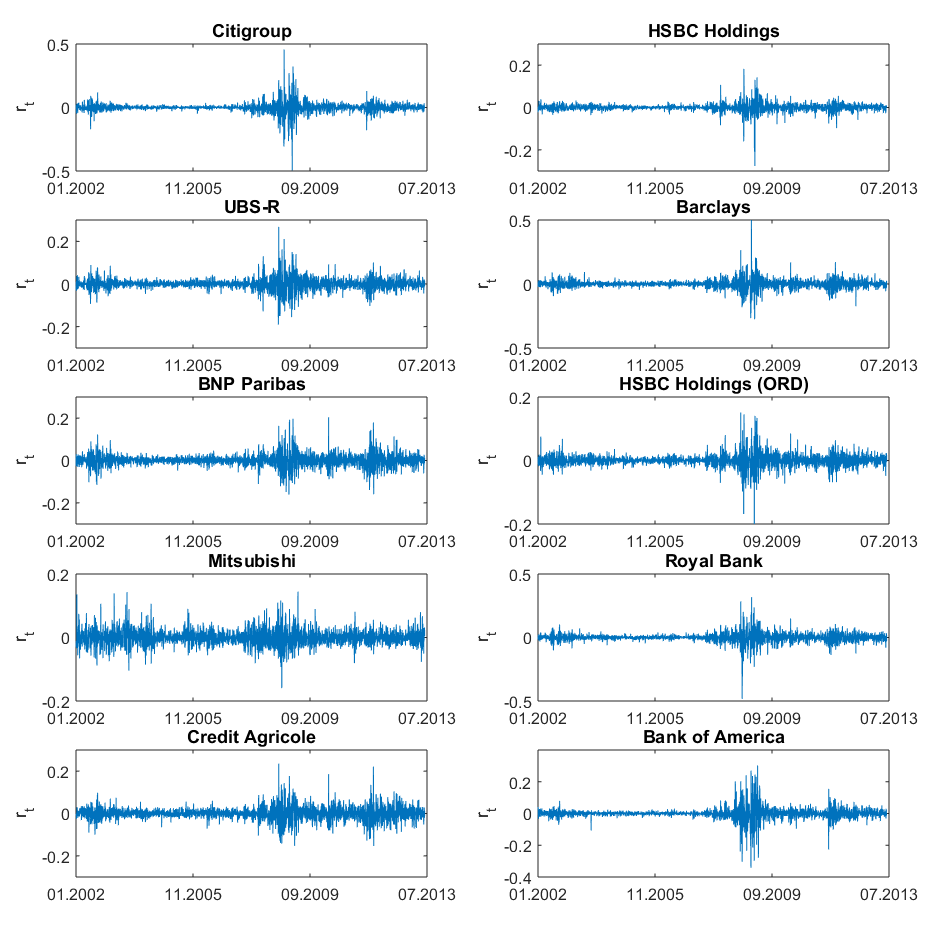}
\caption{Daily log-returns  between 29.01.2002 to 01.07.2013.}
\label{portfolio}
\end{figure}

To illustrate the applicability of the proposed test, we consider a data set of asset returns.
We are interested in the estimation of break points using different dependence measure settings and whether they belong to the same break event.
For a better comparison of similar break dates we use the confidence interval procedure presented in Section \ref{sec:breakpointestimatiion}.
We use daily stock log-returns of ten large firms between 29.01.2002 and 01.07.2013, with a sample size of $T=2980$ and a cross-sectional dimension of $N=10$.
For all returns the closing prices are used.
Table \ref{tab:basistats} presents some basic statistics on the data set.
Plots of the log-returns are presented in Figure \ref{portfolio}, which yield strong fluctuations between 2002-2003, 2007-2008 and 2011-2012 in nearly all assets, indicating a joint behavior during these periods.

\begin{table}[tbp]
\caption{Some basic statistics on the data set of daily log-returns}
\footnotesize
\begin{tabular}{l|rrrrrccc}
 & \multicolumn{1}{c}{Mean} & \multicolumn{1}{c}{SD} & \multicolumn{1}{c}{Skewn.} & \multicolumn{1}{c}{Kurt.} & \multicolumn{1}{c}{AR} & \multicolumn{1}{c}{LB-Test} & \multicolumn{1}{c}{Lag} & \multicolumn{1}{c}{LM-Test} \\ \hline
Citigroup & $-$0.0007 & 0.0364 & $-$0.4967 & 37.2239 & 0.0499 & 0.0000 & 3 & 0.0000 \\ 
HSBC Holdings & 0.0000 & 0.0171 & $-$1.6371 & 47.0280 & $-$0.0461 & 0.0382 & 1 & 0.0000 \\ 
UBS & 0.0000 & 0.0261 & 0.2372 & 14.0963 & 0.0869 & 0.0000 & 3 & 0.0000 \\ 
Barclays & $-$0.0002 & 0.0341 & 1.2502 & 36.4871 & 0.0466 & 0.0000 & 1 & 0.0000 \\ 
BNP Paribas & 0.0001 & 0.0283 & 0.3312 & 10.3057 & 0.0058 & 0.0000 & 3 & 0.0000 \\ 
HSBC Holdings (ORD) & 0.0001 & 0.0188 & $-$0.2019 & 20.2442 & $-$0.0273 & 0.0000 & 1 & 0.0000 \\ 
Mitsubishi UJF & 0.0000 & 0.0242 & 0.3796 & 6.6866 & 0.0407 & 0.0433 & 1 & 0.0000 \\ 
Royal Bank of Scotland & $-$0.0006 & 0.0344 & $-$0.7673 & 27.8513 & 0.0701 & 0.0003 & 1 & 0.0000 \\ 
Crédit Agricole & $-$0.0001 & 0.0297 & 0.2280 & 9.0673 & 0.0268 & 0.0275 & 1 & 0.0000 \\ 
Bank of America & $-$0.0003 & 0.0338 & $-$0.3304 & 26.6730 & $-$0.0299 & 0.0000 & 5 & 0.0000 \\ \hline
\end{tabular}
\label{tab:basistats}
\parbox{16.5cm}{ \vspace{1ex} 
\scriptsize
Note: Some basic statistics on the data set consisting of the log-returns of the shares of ten financial firms between 29.01.2002 and 01.07.2013 are presented. The first four columns correspond to the sample mean, sample standard deviation, sample skewness, and sample kurtosis. The fifth column is the first order sample autocorrelation, and the sixth column corresponds to the p-value of the Ljung-Box test using a maximum lag length of 10 (see \citealp{ljung1978}). The seventh column is the lag length of an AR process for which the BIC is minimized, and the eighth column is the p-value of the LM test for heteroskedasticity (see \citealp{engle1982})
}
\end{table}

\begin{figure}[t]
\centering
\includegraphics[width=0.75\textwidth]{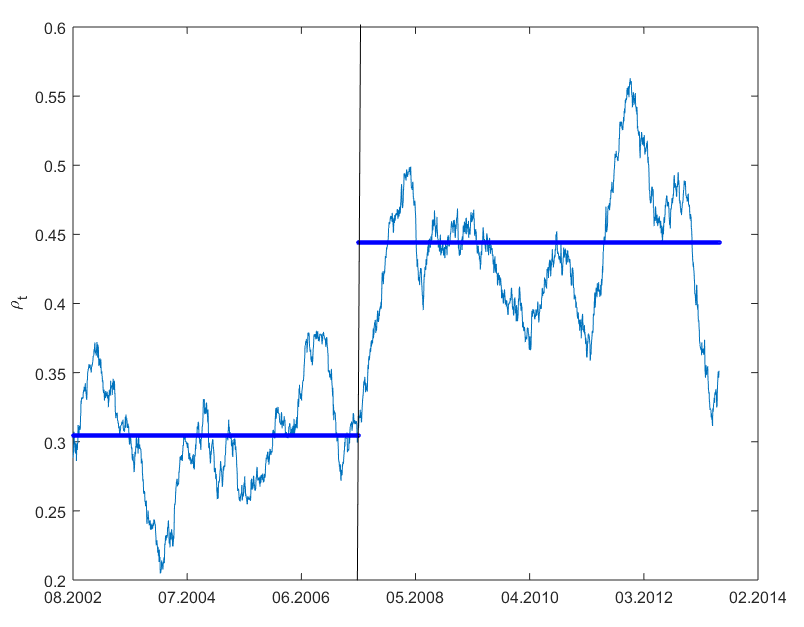}
\caption{Spearman's rank correlation and estimated break point}
\label{rho}
\parbox{12cm}{ \vspace{-0.5ex} 
\scriptsize
Note: The pairwise averaged Spearman's rank correlation coefficients are estimated in a rolling window of size 150.
The vertical line at 17.07.2007 depicts the estimated break point location using approach 1).
The thick blue lines show the estimated rank correlation coefficient from break to break.
}
\end{figure}

To indicate that dependence measures are not time invariant for this data set, we estimate the pairwise averaged Spearman's rank correlation coefficient in a rolling window of size 150, which corresponds to roughly $5\%$ of the sample size.
The plot is depicted in Figure \ref{rho}.
The strongest common fluctuations and high correlation coefficients occur in the period of the peaks of the last financial crisis between early 2007 and late 2008.
The other strong fluctuations and the increase in the correlation can be explained by the decline in stock prices on the stock exchanges in the United States, Canada, Asia, and Europe in October 2002 and the peak of the euro crisis in 2011.

Due to the fact that the test is based on residual data, we first have to estimate a model for each return series $Y_{it}$, $i=1,\dots,N$. 
Table \ref{tab:basistats} shows that the p-values of the Ljung-Box test are smaller than the $5\%$ level, which indicates significant autocorrelation in the data, and the BIC criterion suggests to select a lag length of 1.
Moreover, the LM test by \cite{engle1982} is rejected, which indicates significant heteroskedasticity.
Therefore we follow \cite{oh2013} and use an AR(1)-GARCH(1,1) process to model the conditional mean and variance
\begin{align*}
Y_{i,t}&=\alpha_i+\beta_i Y_{i,t-1}+\sigma_{i,t}\eta_{i,t},\\
\sigma_{i,t}^2&=\gamma_{i,0}+\gamma_{i,1} \sigma_{i,t-1}^2+\gamma_{i,2} \sigma_{i,t-1}^2\eta_{i,t-1}^2,
\end{align*} 
for $t=2,\dots,2980$, where $\eta_{i,t}$ is white noise.

The residuals are then given by $\hat \eta_{i,t} = \hat \sigma_{i,t}^{-1} (Y_{i,t} - \hat \alpha_i - \hat \beta_i Y_{i,t-1})$, where $\hat \alpha_i$, $\hat \beta_i$, and $\hat \sigma_{i,t}$ are the maximum likelihood estimators of the model parameters.
We consider the five dependence vector settings from the simulation section and two approaches for testing and dating break points.
While the first approach can only detect the most significant break point, the second approach can be applied to identify multiple breaks.

\begin{itemize}
\item[1)] We apply the test to the pre-determined residual data considering all information from $t=1$ until $t=T$.
\item[2)] We apply the test in a rolling window setting where we consider periods of size $L$.
If a break point is detected in the period $[t_1,(t_1-1)+L]$ we estimate the break point location $\hat{k}$, and the next considered period is $[t_1+1,t_1+L]$, where $t_1=\hat{k}$.
If no break point is detected, we consider the next time step $t_1+1$. 
The procedure starts at $t_1=1$ and is terminated if $t_1+L>T$.
The marginal models are re-estimated for each period.
\end{itemize}

\begin{table}[t]\caption{List of detected break points using approach 1)}\label{breaks_ap1}
\footnotesize
  \begin{center}
    \begin{tabular}{r||r|r|r|r|r|r}
    \toprule
     & $\hat{m}_1^{ij}$&$\hat{m}_2^{ij}$&$\hat{m}_3^{ij}$&$\hat{m}_4^{ij}$&$\hat{m}_5^{ij}$&avg\\
    \midrule
             $\hat{s}_aT$ & 09.07.2007 &09.07.2007&08.08.2007&09.07.2007&09.07.2007&17.07.2007\\
   \midrule
            $\hat{K}_a^-$& 21.12.2006 &31.08.2006&07.03.2006&30.03.2006&22.02.2007&\\
            $\hat{K}_a^+$& 25.10.2007 &22.11.2007&20.06.2008&28.02.2008&18.10.2007&\\
    \bottomrule
    \end{tabular}
    \end{center} 
    \parbox{13.5cm}{ \vspace{-1.5ex} 
\scriptsize
Note: The detected break points and confidence intervals $[\hat{K}_a^-$,$\hat{K}_a^+]$ for the five dependence vector setting with in the test using approach 1) as well as the average across all break point locations are presented.
}
\end{table}%

\begin{figure}[t]
\centering
\includegraphics[width=0.75\textwidth]{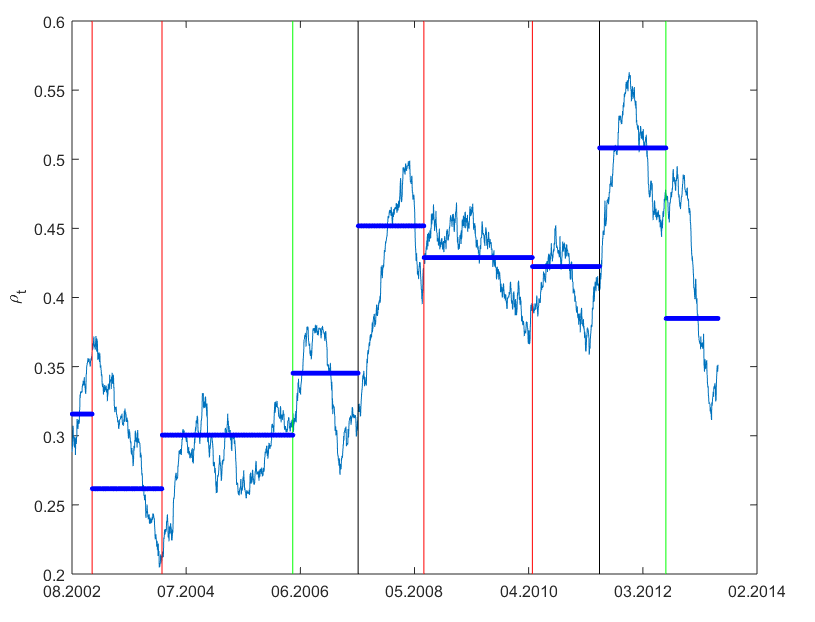}
\caption{Spearman's rank correlation and estimated break points from approach 2)}
\label{rho_all}
\parbox{12cm}{ \vspace{-0.5ex} 
\scriptsize
Note: The pairwise averaged Spearman's rank correlation coefficients are estimated in a rolling window of size 150.
The black vertical lines indicate break point locations averaged over five similar detected points and the red vertical lines indicate break point locations averaged over four similar break points.
Single detected break points are depicted as vertical green lines.
The solid blue line shows estimated rank correlation coefficients from break to break.
}
\end{figure}

The break detection results for approach 1) are shown in Table \ref{breaks_ap1} and Figure \ref{rho}.
The results for approach 2) are presented in Table \ref{breaks_ap2} and Figure \ref{rho_all}.
Using approach 1), almost all dependence settings detect the same break at 09.07.2007, which corresponds to $\hat{s}_1=0.476$.
Only the setting in which the upper quantiles are selected detect a break at 08.08.2007, which corresponds to $\hat{s}_3=0.484$.
However, using the confidence interval procedure from Section \ref{sec:simulations} with the settings $\hat{m}_1^{ij}$ and $\hat{m}_3^{ij}$, we find that the two estimated break point locations $\hat{s}_1=0.476$ and $\hat{s}_3=0.484$ lie in the intersection of both confidence intervals.
Thus, we can conclude that the estimated break point locations belong to the same break event.

While approach 1) gives us the most significant break in our data set, there might be more break point locations.
Figure \ref{rho} reveals that there is a strong increase in the rank correlation coefficient after the break (indicated by the black line) from 0.31 up to 0.44 (indicated by the solid blue line), while the overall maximum change of the rolling window estimates is even higher between 0.27 and 0.49.

\begin{table}[t]\caption{List of detected break points using approach 2)}\label{breaks_ap2}
\footnotesize
  \begin{center}
    \begin{tabular}{r||r|r|r|r|r|r}
    \toprule
     year& $\hat{m}_1^{ij}$&$\hat{m}_2^{ij}$&$\hat{m}_3^{ij}$&$\hat{m}_4^{ij}$&$\hat{m}_5^{ij}$&avg\\
    \midrule

  2002/2003   &        30.12.2002    &08.01.2003&20.12.2002&          &23.12.2002&27.12.2002\\
       2004   &        19.02.2004    &26.02.2004&          &05.03.2004&04.03.2004&01.03.2004\\
       2005   &                      &          &          &          &          &\\
       2006   &                      &          &          &11.05.2006&          &11.05.2006\\
       2007   &        24.07.2007    &11.07.2007&17.07.2007&16.02.2007&09.07.2007&15.06.2007\\
       2008   &        16.07.2008    &08.08.2008&16.07.2008&17.07.2008&          &23.07.2008\\
       2009   &                      &          &          &          &          &\\
       2010   &        21.04.2010    &15.06.2010&29.04.2010&10.06.2010&          &19.05.2010\\
       2011   &        28.06.2011    &28.06.2011&21.09.2011&14.06.2011&20.05.2011&05.07.2011\\
       2012   &                      &          &          &14.08.2012&          &15.08.2012\\
    \bottomrule
    \end{tabular}
    \end{center} 
\parbox{14.5cm}{ \vspace{-1.5ex} 
\scriptsize
Note: The detected break points for the five dependence vector setting under approach 2) as well as the average across all break point locations for each break event are reported.
}
\end{table}%

For approach 2), we consider a rolling window of size $L=400$, which corresponds to roughly 15 percent of the sample size.
Note that there is multiple testing when using approach 2).
This should be kept in mind when interpreting test results and confidence levels should be adjusted accordingly.
Most detections in Table \ref{breaks_ap2} can be explained by well known financial market crashes from the last twenty years. 
Breaks are detected during the fall in stock prices on the stock exchanges in the United States, Canada, Asia, and Europe in October 2002, 
the start of the Iraq war in 2003/2004, the beginning of the last financial crisis in 2007, the bankruptcy of Lehman Brother's in 2008, and the euro crisis, starting at the end of 2009, with its peak in 2011.
The results indicate that the break point estimates of the dependence settings $\hat{m}_1^{ij}, \hat{m}_2^{ij}$, and $\hat{m}_3^{ij}$ seem to be closely related and belong to the above mentioned events.
The break event in 2004 is not significant for the upper quantile setting. 
In contrast to the simulation study, where the setting $\hat{m}_5^{ij}$ has the highest power, the test for this setting detects only four significant break dates at 2002, 2004, 2007 and 2011, which are overall the most significant.
For the lower quantile setting $\hat{m}_4^{ij}$, we obtain a slightly different result, where the tested periods in 2002/2003 are not significant. On the other hand, breaks are detected in the mid of 2006 and 2012.

An advantage of using different dependency settings is that we can conclude that a detected break point in certain regions of the distribution is in some ways more relevant if it is detected by more than one setting.
The break events in the years 2002, 2004, 2008, and 2010 are detected by four settings, and the break events in the years 2007 and 2011 even by five settings.
These are the most significant breaks and can be explained by well known financial market crashes, as mentioned above.
The break point detected in 2007 is also in line with the break event detected when using approach 1), which corresponds to the highest dependence change in the test period considered.
When using the setting $\hat{m}_4^{ij}$, we obtain different estimated break points and indicates that flexible dependence measure settings might yield a clearer and wider interpretation of the results.
Since most of the detected break dates correspond to the same break events and are in most cases very close to each other, we average the break dates over all settings (see the last column in Table \ref{breaks_ap2}).

\begin{figure}[t]
\centering
\includegraphics[width=0.75\textwidth]{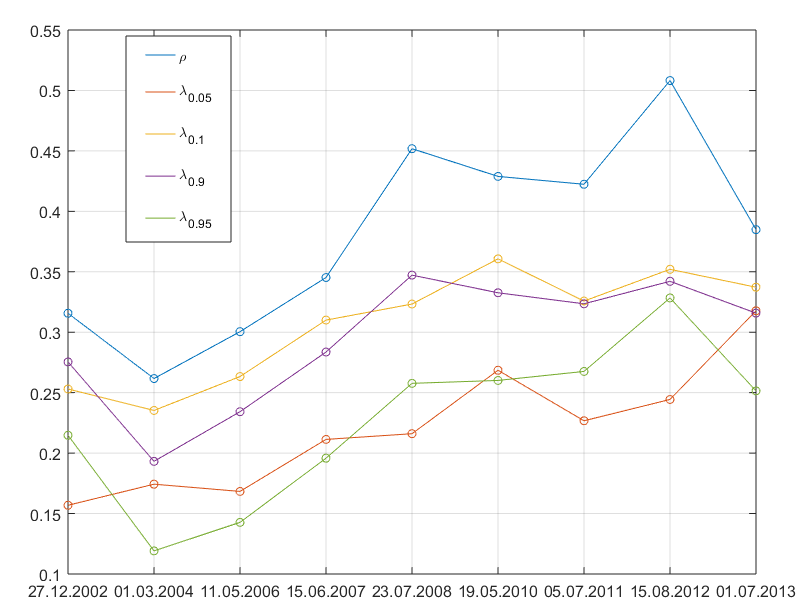}
\caption{Spearman's rank correlation and quantile dependencies}
\label{depend}
\parbox{11.5cm}{ \vspace{0.5ex} 
\scriptsize
Note: Spearman's rank correlation and $(0.05, 0.1, 0.9, 0.95)$-quantile dependencies are reported. The dependence measures are estimated from break to break.
}
\end{figure}

In Figure \ref{rho_all}, the pairwise averaged Spearman's rank correlation coefficients in a rolling window are plotted, and the averaged break point estimates from approach 2) together with the rank correlation estimates from break to break are depicted.
The high jumps in the rank correlation are noticeable in the periods of the most significant breaks in 2007 and 2011, where, if break to break estimates are considered, the correlation jumps from 0.35 to 0.45 (2007) and 0.42 to 0.51 (2011). 
Furthermore the upper quantile dependence measures increase strongly (see Figure \ref{depend}).

In general, nearly all detected break events correspond to an increase of the considered dependence measures (see Figure \ref{depend}), except for the first detected break in 2002/2003, where the peak in this period of nearly all dependencies is reached. 
We also see a decrease in the rank correlation and 0.9-quantile dependence in the period after the last financial crisis.
Overall, we observe an increase in the dependencies among the asset returns. 
Finally, we are interested in the break behavior if we divide our considered portfolio in subsets of assets. 
We split our portfolio in two groups of five assets, where we collected the assets with the highest unconditional variance in the first group.
For the group with the highest variance we consider the log-returns of Citigroup, Barclays, Royal Bank of Scotland, Crédit Agricole, and Bank of America (group 1), where the other group consists of HSBC Holdings, UBS, BNP Paribas, HSBC Holdings (ORD), and Mitsubishi UJF (group 2).
We apply the test separately for each group of assets and consider all five dependence vector settings using approach 1). 
Table \ref{breaks_sub} shows the estimated break point locations and confidence intervals for group 1 (upper panel) and group 2 (lower panel).

\begin{table}[t]\caption{Detected break points for group 1 and 2}\label{breaks_sub}
\footnotesize
  \begin{center}
    \begin{tabular}{r||r|r|r|r|r}
    \toprule
    Group 1 & $\hat{m}_1^{ij}$&$\hat{m}_2^{ij}$&$\hat{m}_3^{ij}$&$\hat{m}_4^{ij}$&$\hat{m}_5^{ij}$\\
    \midrule
             $\hat{s}_aT$ & 09.07.2007 &16.02.2007&26.11.2007&16.02.2007&09.07.2007\\
   \midrule
            $\hat{K}_a^-$& 24.11.2006 &22.02.2006&09.12.2005&23.12.2005&14.03.2007\\
            $\hat{K}_a^+$& 27.09.2007 &25.05.2007&05.12.2008&06.07.2007&26.09.2007\\
    \bottomrule
     \toprule
    Group 2 & $\hat{m}_1^{ij}$&$\hat{m}_2^{ij}$&$\hat{m}_3^{ij}$&$\hat{m}_4^{ij}$&$\hat{m}_5^{ij}$\\
    \midrule
             $\hat{s}_aT$ & 09.07.2007 &09.07.2007&&&31.05.2006\\
   \midrule
            $\hat{K}_a^-$& 23.11.2005 &18.02.2005&&&13.07.2004\\
            $\hat{K}_a^+$& 19.06.2008 &10.10.2008&&&14.03.2007\\
    \bottomrule
    \end{tabular}
    \end{center} 
 \parbox{12cm}{ \vspace{-1.5ex} 
\scriptsize
Note: Detected break points and corresponding confidence intervals $[\hat{K}_a^-$,$\hat{K}_a^+]$ for Group 1 (upper panel) and Group 2 (lower panel) under approach 1) are reported.
}
\end{table}

Using the first group data, all dependence measure settings detect a significant break at the $0.05$ significance level, whereas in the second group only the breaks of the settings $\hat{m}^{ij}_1, \hat{m}^{ij}_2$ and $\hat{m}^{ij}_5$ are significant. 
The results are mixed compared to Table \ref{breaks_ap1}. 
Most of the detected breaks correspond to the summer of 2007, and the same break date is detected at 09.07.2007 using the settings $\hat{m}^{ij}_1$, $\hat{m}^{ij}_5$ for group 1, and $\hat{m}^{ij}_1$, $\hat{m}^{ij}_2$ for group 2. 
When considering the first group data, the settings $\hat{m}^{ij}_2$ and $\hat{m}^{ij}_4$ find an earlier break at 16.02.2007, while the setting $\hat{m}^{ij}_3$ detects a break at 26.11.2007. 
A more distinct break is detected within the second group data at 31.05.2006 using the setting $\hat{m}^{ij}_5$, where only Spearman's rank correlation coefficient is used. 
Using the common break procedure, we find that the break event at 31.05.2006 is clearly distinct from the break event at 09.07.2007. 
The found break in 2006 may correspond to the early beginning of the last financial crisis in the summer of 2006. 
Further, the break event at 16.02.2007 is also distinct to the one at the mid of 2007, whereas this separation is not as pronounced as the previous one.

\section{Conclusion}\label{sec:conclusion}

We apply the test by \cite{manner2019}, which is designed for testing breaks in factor copula models, to the more general problem of testing for changes in cross-sectional dependencies of multivariate time series.
The asymptotic null distribution is not known in closed form and therefore estimated by an i.i.d.\ bootstrap procedure.
A bootstrap procedure to date break points and to decide whether two detected break points are driven by the same date event is discussed, and a comprehensive simulation study is provided.

A size and power analysis, using different dependence measure settings for different simulated fat and skewed distributed data shows that the best power properties are obtained when considering solely Spearman's rank correlation and a combination of Spearman's rank correlation and quantile dependencies, where the simple setting when using only the rank correlation coefficient works best.
The settings when using only upper or lower quantile dependencies have poor power properties. 
Furthermore, we found that using upper quantile dependencies results in better power properties in the presence of strongly left skewed data compared to lower quantile dependencies, and, on the other hand, lower quantile dependencies result in better power properties when considering right skewed data compared to the usage of upper quantile dependencies.
Considering lower and upper quantile dependencies jointly results for all cases in better power properties than when using these separately.
The test is also applied to real data to indicate the usefulness of choosing different dependence measure settings. 
We consider historical data of daily log-returns of ten large companies during the last financial crisis from 2002 to the mid of 2013. 
One advantage of using different dependence settings is that we can compare the detected breakpoints from different settings and draw conclusions about whether certain breakpoints are more pronounced than others.
On the other hand, we also get a different picture of the break points, which motivates the use of flexible dependence measure settings and the combination of rank correlations and quantile dependencies. 
Furthermore, we propose a heuristic procedure to be able to make a statement about the equality of two estimated break point locations.

\newpage
\bibliographystyle{ecta}
\bibliography{references}

\end{document}